\documentclass[%
reprint,
showpacs,
bibnotes,
amsmath,amssymb,
aps,
prl,
floatfix,
]{revtex4-1}

\usepackage{graphicx}
\usepackage{dcolumn}
\usepackage{bm}
\usepackage{hyperref}
\usepackage{multirow}
\usepackage{array}
\usepackage{booktabs}
\usepackage{ctable}
\usepackage{upgreek}
\usepackage{epsfig}
\usepackage{mathrsfs}
\usepackage{amssymb}
\usepackage{amsbsy}
\usepackage{color}
\usepackage{cancel}
\usepackage{marginnote}
\usepackage{amsfonts}
\newcommand{\bcen}{\begin{center}}
\newcommand{\ecen}{\end{center}}
\newcommand{\btab}{\begin{tabular}}
\newcommand{\etab}{\end{tabular}}
\newcommand{\bdes}{\begin{description}}
\newcommand{\edes}{\end{description}}

\newcommand{\ul}{\underline}
\newcommand{\beq}{\begin{equation}}
\newcommand{\eeq}{\end{equation}}
\newcommand{\bea}{\begin{eqnarray}}
\newcommand{\eea}{\end{eqnarray}}

\newcommand{\half}{\frac{1}{2}}
\newcommand{\bary}{\begin{array}}
\newcommand{\eary}{\end{array}}
\newcommand{\benum}{\begin{enumerate}}
\newcommand{\eenum}{\end{enumerate}}
\newcommand{\bitem}{\begin{itemize}}
\newcommand{\eitem}{\end{itemize}}

\newcommand{\bsig}{\mbox{\boldmath $ \sigma $}}

\newcommand{\bOne}{{\boldsymbol 1}}
\newcommand{\be} { \mbox{\boldmath $e$}}

\newcommand{\bk} { \bm{k} }

\newcommand{\bq} { \bm{q} }
\newcommand{\br} { \boldsymbol{r}}

\newcommand{\dH}{{\mathbb{H}}}

\newcommand{\D}[1]{\mbox{d}{#1}}

\newcommand{\mean}[1]{\langle #1 \rangle}

\newcommand{\ket}[1]{| #1 \rangle}
\newcommand{\braket}[2]{\langle #1 | #2 \rangle}

\newcommand{\detune}{{\updelta}}
\newcommand{\eqn}[1] {eqn.~(\ref{#1})}

\newcommand{\Fig}[1]{Fig.~\ref{#1}}

\makeatletter

\newcommand{\Rmnum}[1]{\expandafter\@slowromancap\romannumeral #1@}
\makeatother

\newlength{\myfigwidth}
\newlength{\myhalffigwidth}
\newcommand{\asc}{a_{sc}}
\newcommand{\as}{a_{s}}

\newcommand{\kf}{k_F}

\newcommand{\halfsum}[1]{\displaystyle{\sum^{\sim}_{#1}}}

\newcommand{\sthre}{{\varepsilon_{th}^s}}
\newcommand{\ethre}{{\varepsilon_{th}}}
\newcommand{\ebse}{{\varepsilon_{bs}}}
\newcommand{\egse}{{\varepsilon_{gs}}}

\newcommand{\mylabel}[1]{\label{#1}} 

\newcommand{\mycite}[1]{\cite{#1}}

\usepackage{lineno}

\begin{document}



\title{Flow enhanced pairing and other novel effects in Fermi gases in synthetic gauge fields}

\author{Vijay B. Shenoy}
\email{shenoy@physics.iisc.ernet.in}

\affiliation{Centre for Condensed Matter Theory, Department of Physics, Indian Institute of Science, Bangalore 560 012, India}



\date{\today}

\begin{abstract}
Recent experiments on fermions in synthetic gauge fields result in systems with a spin-orbit coupling along one spatial axis, a detuning field, and a Zeeman field. We show theoretically that the presence of {\em all three} results in interesting and unusual phenomena in such systems in the presence of a contact singlet attraction  between the fermions (described by a scattering length). For two particles, bound states appear over certain range of the centre of mass momenta when a critical positive scattering length is attained, with the deepest bound state appearing at a {\em nonzero} centre of mass momentum.  For the centre of mass momenta without a bound state,  the gauge field induces a resonance like feature in the scattering continuum resulting in a large scattering phase shift.  For many particles, we demonstrate that the system, in a parameter range, shows flow enhanced pairing, i.~e., a more robust superfluid at finite centre of mass momentum. Yet another regime of parameters offers the opportunity to study strongly interacting normal states of spin-orbit coupled fermionic systems utilizing the resonance like feature induced by the synthetic gauge field.
\end{abstract}


\maketitle

Cold atomic systems with synthetic gauge
fields\mycite{Dalibard2011} are expected to greatly enhance their ability to simulate exotic phases of matter from
fractional quantum hall states to strongly coupled gauge
theories. Spectacular experimental advances\mycite{Lin2009A,
  Lin2009B, Lin2011} have been successful in creating and
studying systems with synthetic gauge fields.

These advances have resulted in a flurry of theoretical
activity directed towards understanding interacting fermions in synthetic gauge
fields. Uniform non Abelian gauge fields that produce a generalized
Rasbha spin-orbit coupling have been investigated. Even with a weak attractive
interaction, increasing the magnitude of such spin orbit
coupling produces a crossover from a BCS state
with large pairs to a BEC like state which is a condensate of
new type of boson called the
rashbon.\mycite{Vyasanakere2011BCSBEC} Superfluidity of
rashbons\mycite{Hu2011}, BCS-BEC crossover with Zeeman
fields etc\mycite{Gong2011,Iskin2011a,Han2011}, and transition
temperatures of the rashbon condensate\mycite{Yu2011,Vyasanakere2011Rashbon}, have been studied (see
\mycite{Shenoy2012CurSci}, for a review).

Very recent experimental studies on fermions with synthetic gauge fields by the Shanxi\mycite{Shanxi2012} and MIT\mycite{MIT2012} groups have produced systems with a spin orbit coupling ($\lambda$) in one spatial direction (equal mixture of Rashba and Dresselhaus), a detuning ($\detune$) field,  and a Zeeman ($\Omega$) field. In this paper, we explore the rich physics of such systems by theoretical studies of interacting fermions in them. We show that the {\em simultaneous} presence of the detuning and Zeeman terms ($\detune \ne 0, \Omega \ne 0$), together with $\lambda$, produces several interesting and unusual effects. At the two body level, a bound state exists only for a range of the centre of mass momenta, and the most strongly bound state (largest binding energy) appears at a {\em finite centre of mass momentum}. For centre of mass momenta where the bound state is absent, a resonance like feature with associated large change in the scattering phase shift appears in the {\em scattering continuum}. Interesting consequences that obtain in the many body setting include flow enhanced pairing (larger pairing amplitude for a flowing superfluid!). Our work also clearly bring out the possibility of studying strongly interacting normal states of spin-orbit coupled fermions engendered by resonance like feature in the scattering continuum induced by the gauge field.

The non-interacting part of the Hamiltonian of the system of interest is given by 
\beq\mylabel{eqn:RSO}
{\cal H} = \sum_{\bk, \sigma \sigma'} c^\dagger_{\bk \sigma} \left[\frac{{\bk}^2}{2} \delta_{\sigma \sigma'} -( \lambda k_x -\detune) \tau^x_{\sigma \sigma'} - \Omega \tau^z_{\sigma \sigma'} \right] c_{\bk \sigma'},
\eeq
where $c$s are fermion operators, $\detune$ and $\Omega$ are the detuning and two-photon Rabi coupling (Zeeman field) respectively, $\tau$s are Pauli matrices. As noted above, this type of spin-orbit coupling\mycite{Liu2009} has been realized in recent experiments.\mycite{Lin2011,Shanxi2012,MIT2012} The one-particle states of ${\cal H}$ have energies
\beq\mylabel{eqn:Dispersion}
\varepsilon_\alpha(\bk) = \frac{k^2}{2} - \alpha B(\bk), \;\;\; B(\bk) = \sqrt{(\lambda k_x - \detune)^2 + \Omega^2},
\eeq
where $\alpha = \pm 1$ is the generalized helicity (simply called helicity below), with the associated states 
\beq\mylabel{eqn:1PStates}
\begin{split}
\ket{\bk\alpha} &= \ket{\bk}\otimes \ket{\chi_\alpha(\bk)}, \;\;\;  \ket{\chi_\alpha(\bk)} = \sum_\sigma f^\alpha_\sigma(\bk) \ket{\sigma},  \\
 f^\alpha_{\sigma}(\bk) &= \frac{1}{\sqrt{2}} \left(\cos{\theta_\alpha(\bk)} + \sigma \sin{\theta_\alpha(\bk)} \right),
\end{split}
\eeq
where $\theta_\alpha(k) = \arctan\left(\frac{\Omega}{\lambda k_x - \detune} \right) + (1 - \alpha) \frac{\pi}{2}$.

We introduce a contact singlet attraction between the spin-$\half$ fermions (such as that in a broad Feshbach resonance\cite{Chin2010})
\beq
{\cal H}_{\upsilon} = \frac{\upsilon}{2} \int \D{^3 \br} \, S^\dagger(\br) S(\br) = \frac{\upsilon}{2 V}\sum_{\bq} S^\dagger(\bq) S(\bq),
\eeq
where $S^\dagger(\bq) = \sum_{\bk}  \frac{1}{\sqrt{2}} \left( c^\dagger_{\frac{\bq}{2} + \bk \uparrow} c^\dagger_{\frac{\bq}{2} - \bk \downarrow}  -  c^\dagger_{\frac{\bq}{2} + \bk \downarrow} c^\dagger_{\frac{\bq}{2} - \bk \uparrow}  \right)$,
and the bare parameter $\upsilon$ is related to the $s$-wave scattering length $\as$ through an ultraviolet cutoff $\Lambda$ as $\frac{1}{4 \pi \as} = \frac{1}{\upsilon} + \Lambda$.
We  the study the ground state of a finite density of fermions in this system, after an investigation of the two-body problem.

\noindent
{\sc{Two-Body Problem:}}  It is convenient to introduce the following two body states: the singlet state
\beq\mylabel{eqn:Singlet}
\ket{\bq,\bk,s} = S^\dagger(\bq,\bk) \ket{0}
\eeq
and the state
\beq\mylabel{eqn:TwoPartile}
\ket{\bq,\bk,\alpha \beta} = c^\dagger_{\frac{\bq}{2} + \bk\alpha} c
^\dagger_{\frac{\bq}{2} - \bk\beta} \ket{0},
\eeq
where $\ket{0}$ is the fermion vacuum. We define the singlet amplitude as
\beq\mylabel{eqn:SingletAmplitude}
A_{\alpha \beta}(\bq,\bk) = \braket{\bq,\bk,s}{\bq,\bk,\alpha \beta},
\eeq
which for our system is
\beq
|A_{\alpha \beta}(\bq,\bk)|^2 = \frac{1}{4} \left[1 - \alpha \beta \left( \frac{(\lambda k_x^+ - \detune) (\lambda k_x^- - \detune) + \Omega^2 }{B(\bk^+)B(\bk^-) } \right) \right],
\eeq
where $\bk^\pm = \frac{\bq}{2} \pm \bk$.

\begin{figure}
\centerline{\includegraphics[width=\myfigwidth]{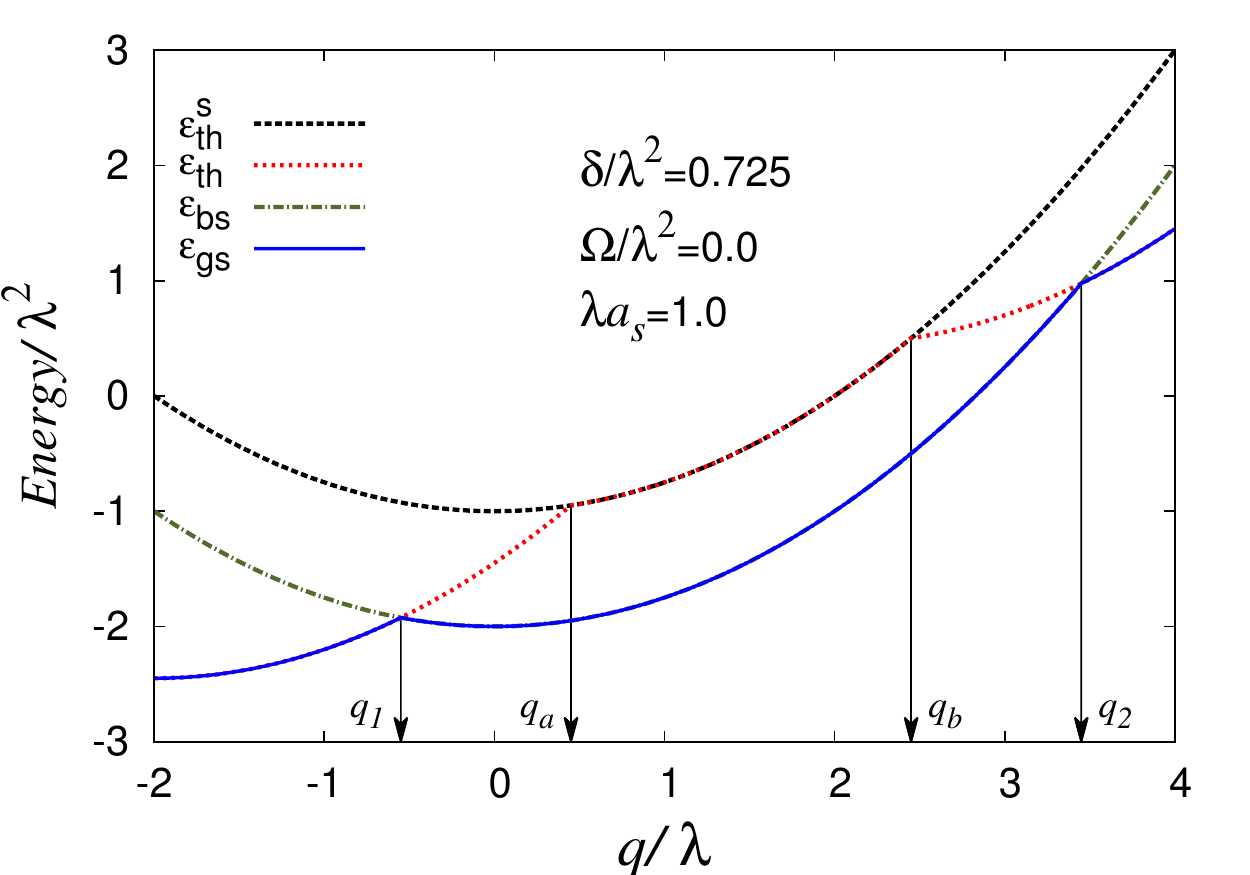}}
\caption{{\bf Two body problem with no Zeeman field ($\detune \ne 0, \Omega=0$): }Singlet threshold $\sthre$, two-body threshold $\ethre$, bound state energy $\ebse$, and ground state energy $\egse$ as a function of $q$. }
\mylabel{fig:Detune}
\end{figure}

Solution of $\omega$ of  the secular equation 
\beq\mylabel{eqn:SecularEquation}
\frac{1}{4 \pi \as} - \frac{1}{V} \sum_{\bk} \left[ \left(\sum_{\alpha \beta} \frac{|A_{\alpha \beta}(\bq,\bk)|^2}{\omega - \varepsilon_{\alpha \beta}(\bq,\bk)} \right) + \frac{1}{k^2} \right] = 0,
\eeq
where $V$ is the volume of the system, and
\beq\mylabel{eqn:epsalphabeta}
\varepsilon_{\alpha \beta}(\bq,\bk) = \varepsilon_\alpha(\frac{\bq}{2} + \bk) + \varepsilon_\beta(\frac{\bq}{2} - \bk),
\eeq
 below the scattering threshold gives the energy of the bound state.  The physics behind bound state formation\mycite{Vyasanakere2011TwoBody} is determined by the singlet density of states (dos)
\beq\mylabel{eqn:SingletDos}
g_s(\bq,\omega) = \frac{1}{V} \sum_{\bk,\alpha \beta} |A_{\alpha \beta}(\bq,\bk)|^2 \delta(\omega - \varepsilon_{\alpha \beta}(\bq,\bk)).
\eeq
The key aspect to be noted is that the singlet scattering threshold, $\sthre(\bq) $, the smallest value of $\omega$ such that $g_s(\bq,\omega)=0^+$, can be above the two-body threshold $\ethre (\bq)= \min_{\bk} \varepsilon_{++}(\bq,\bk)$, i.~e.,
\beq
\ethre(\bq) \le \sthre(\bq).
\eeq
We now discuss the solution of the two body problem in various regimes of parameters. We consider  centre of mass momenta of the type $\bq = q \be_x$ which contain  interesting physics.

\begin{figure}
\centerline{\includegraphics[width=\myfigwidth]{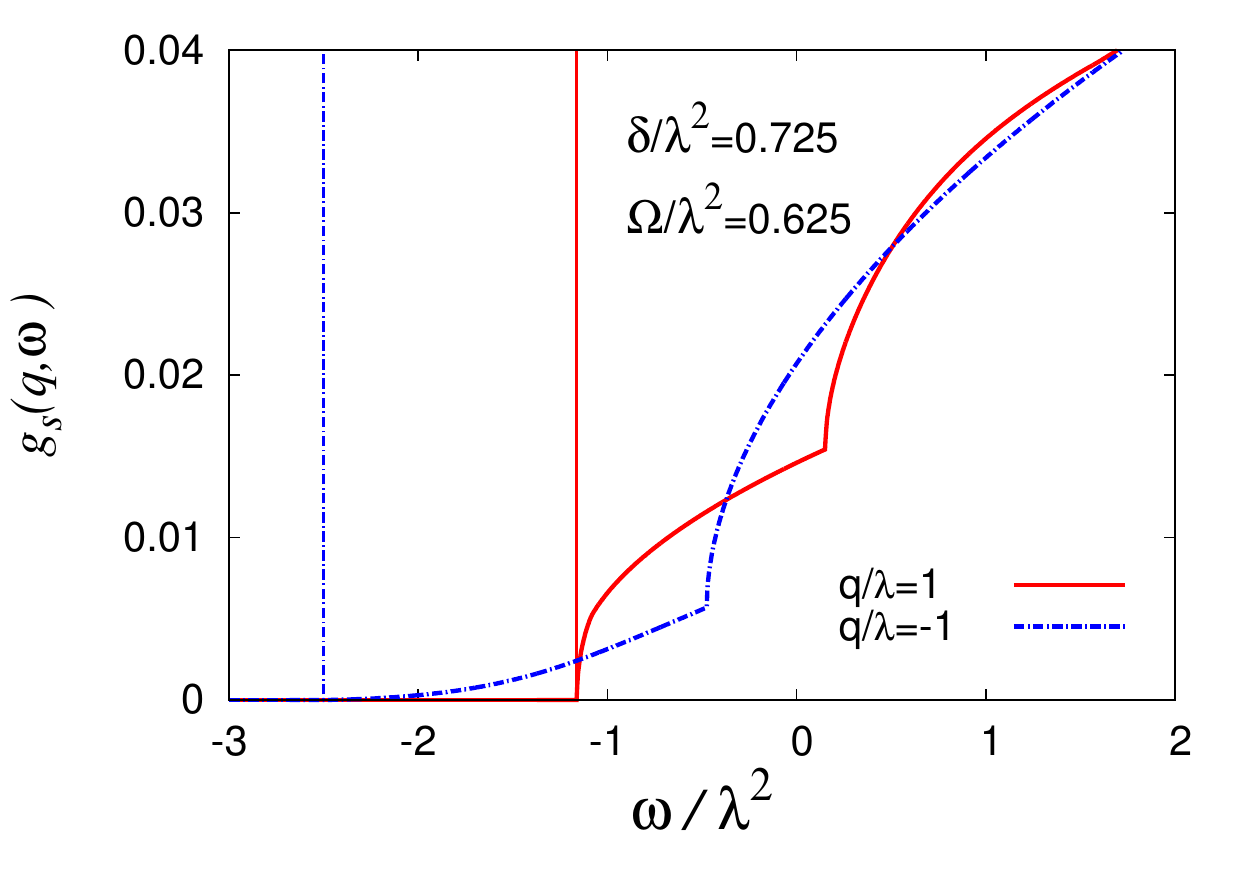}}
\caption{{\bf Singlet dos for $\detune \ne0, \Omega \ne 0$:} The singlet threshold $\sthre$ coincides with the two-body threshold $\ethre$ for all $q$. Plot shows $g_s(q,\omega)$ for two values of $q$, $q= \lambda$ and $q = -\lambda$. The vertical lines correspond to respective threshold energies.}
\mylabel{fig:Sdos}
\end{figure}

\medskip

\noindent
{\bf Case A ($\detune \ne 0,\Omega=0$):} Here we consider the case where $\lambda$ and $\detune$ are non-vanishing.
In the absence of the Zeeman field $\Omega$, we obtain  an analytical expression for the threshold
\beq
\sthre = \frac{q^2}{4} - \lambda^2,
\eeq
with $g_s(\bq,\omega) = \frac{1}{4 \pi^2} \sqrt{\omega - \sthre}$. The interesting aspect of this result is that, despite the presence of $\detune$, which breaks the $k_x \leftrightarrow -k_x$ symmetry, the singlet dos satisfies $g_s(-\bq,\omega) = g_s(\bq,\omega)$.  The singlet threshold coincides with the scattering threshold only in the regime $q_a \le q \le q_b$ (see \Fig{fig:Detune}), where $q_a,q_b$ depend on $\lambda$ and $\detune$. For $q$ outside this interval the scattering threshold $\ethre$ is {\em below} the singlet threshold. This is because the lowest two body state which corresponds to both the particles with $+$ helicity has no singlet component when $q$ is outside the interval $(q_a,q_b)$.

 A scattering length  $\as > 0$, results in a binding energy of $1/\as^2$ and a bound state energy, $\ebse(\bq) = -\frac{1}{\as^2} + \sthre(\bq)$ (see \Fig{fig:Detune}). Note that the energy of this state is satisfies, $\ebse(-\bq) = \ebse(\bq)$, owing to the similar property of the singlet dos.  This bound state is the ground state of the two-body system only in a regime of $q$, i.~e., $q_1 \le q \le q_2$ as shown in \Fig{fig:Detune}, where $q_1, q_2$ depend on $\lambda, \detune$ and $\as$. There is, therefore, a ``first order'' transition in the ground state at $q_1$ (and $q_2$) where the ground state abruptly changes from the bound state with a large singlet component, to a ``free'' triplet state with both particles of $+$ helicity.

\medskip
\noindent
{\bf Case B ($\detune \ne 0, \Omega \ne 0$):} How does a Zeeman field affect the picture just discussed? The key effect of the Zeeman field is on the singlet dos. In the presence of a Zeeman field however small, the singlet threshold $\sthre$ coincides with the two-body threshold $\ethre$ for {\em all} $q$. There is an important additional effect. Although the singlet threshold is reduced, the singlet density of states does {\em not} have the characteristic $\sqrt{\omega - \sthre}$ behaviour for all $q$. For the case shown in \Fig{fig:Sdos}, for $q = \lambda$, the singlet dos has the usual $\sqrt{\omega - \sthre}$ threshold behaviour. In contrast, for $q = -\lambda$ the threshold behaviour is quite different. Analytical calculations show that the singlet dos in this case goes as $(\omega - \sthre)^{3/2}$. It can in fact be shown that this is the case whenever $q$ is outside an interval $(q_a,q_b)$ (\Fig{fig:Omega}). Indeed in \Fig{fig:Sdos}, $q = -\lambda$ corresponds to a $q$ that lies outside the said interval. The gist is that when $\detune\ne 0$ \ul{and} $\Omega \ne 0$, the density of states is not symmetric in $q$, i.~e., $g_s(-q \be_x,\omega) \ne g_s(q \be_x, \omega)$. This is to be contrasted with the previous discussion when $\Omega=0$, where $g_s$ was $q$-symmetric even when $\detune\ne0$.

\begin{figure}
\centerline{\includegraphics[width=\myfigwidth]{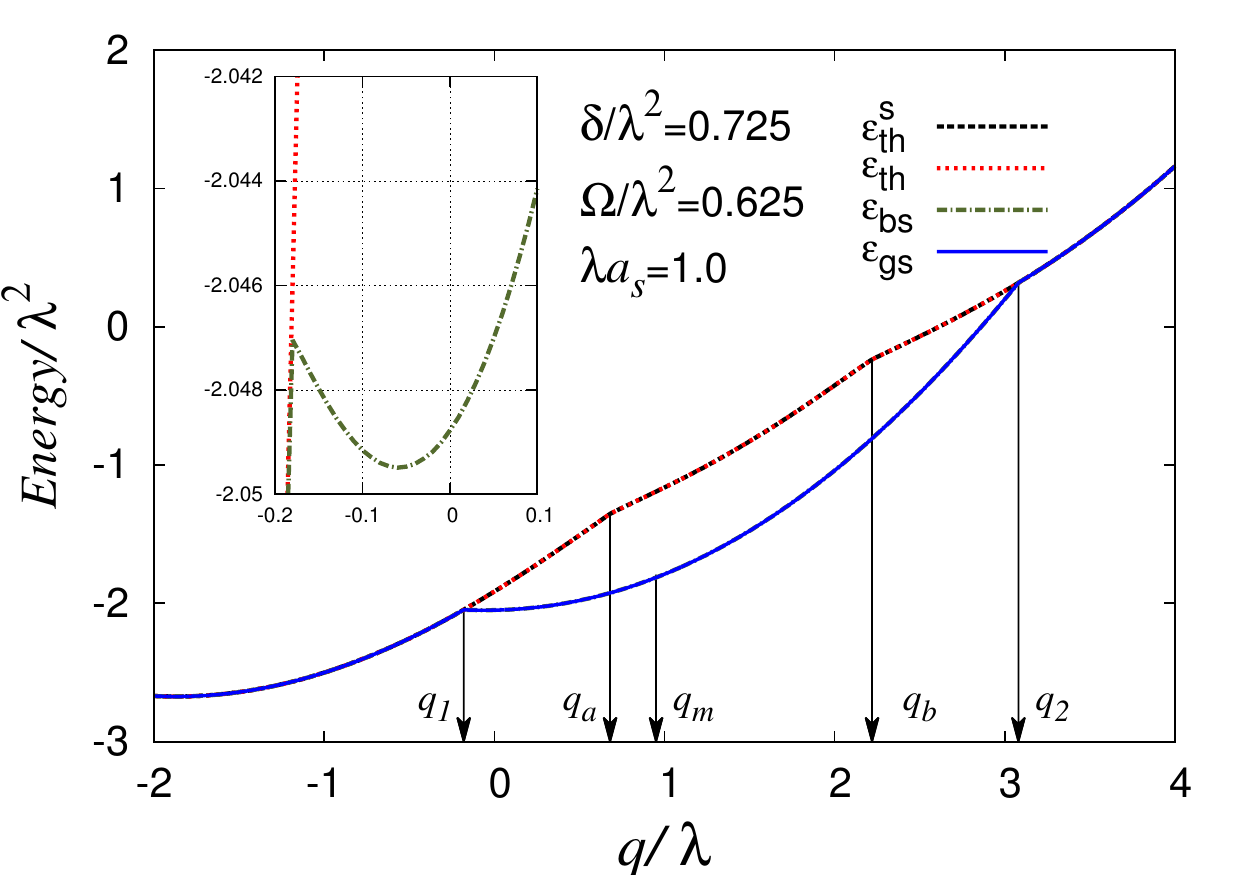}}
\caption{{\bf Two body problem with Zeeman field ($\Omega\ne0$):} Singlet threshold $\sthre$, two-body threshold $\ethre$, bound state energy $\ebse$, and ground state energy $\egse$ as a function of $q$.   In this case the two-body threshold and singlet threshold always coincide. The deepest bound state (largest binding energy) appears at a nonzero value of $q$, at $q_m$. Inset shows the $q$-asymmetry of the bound state energy. }
\mylabel{fig:Omega}
\end{figure}

These characteristics have interesting effects on the ground state of
two particles. A positive critical scattering length is
necessary to produce a bound state, owing to the Zeeman
field. Even for a positive scattering length smaller than this
critical scattering length, as shown in \Fig{fig:Omega}, a well
defined bound state exists only when $q_1 \le q \le q_2$.  The deepest bound state appears at a finite momentum $q_m$ as shown in \Fig{fig:Omega}, and the bound state energy is not at symmetric function of $q$ (\Fig{fig:Omega},inset).

 When $q$ is outside the interval $q_1 \le q \le q_2$ (\Fig{fig:Omega}), bound states cease to exist. However, the ``remanent'' of the bound state appears as a ``resonance like'' feature in the scattering continuum with a large change in phase shift $\eta$ \cite{Taylor2006} appearing at a $q$-dependent energy in the scattering continuum (see \Fig{fig:PhaseShift}). This is akin to what is found in narrow Feshbach resonances (without spin-orbit coupling)\mycite{HoCui2012}. The notable feature here is that this resonance like feature is induced by synthetic gauge field, i.~e., the spin-orbit coupling along with the detuning and Zeeman fields, even when the interaction between the fermions is from a wide resonance described by an energy independent scattering length.  As is evident, this will have profound consequences for the many body problem as strong fermion-fermion interactions can be produced in the scattering continuum.

\begin{figure}
\centerline{\includegraphics[width=\myfigwidth]{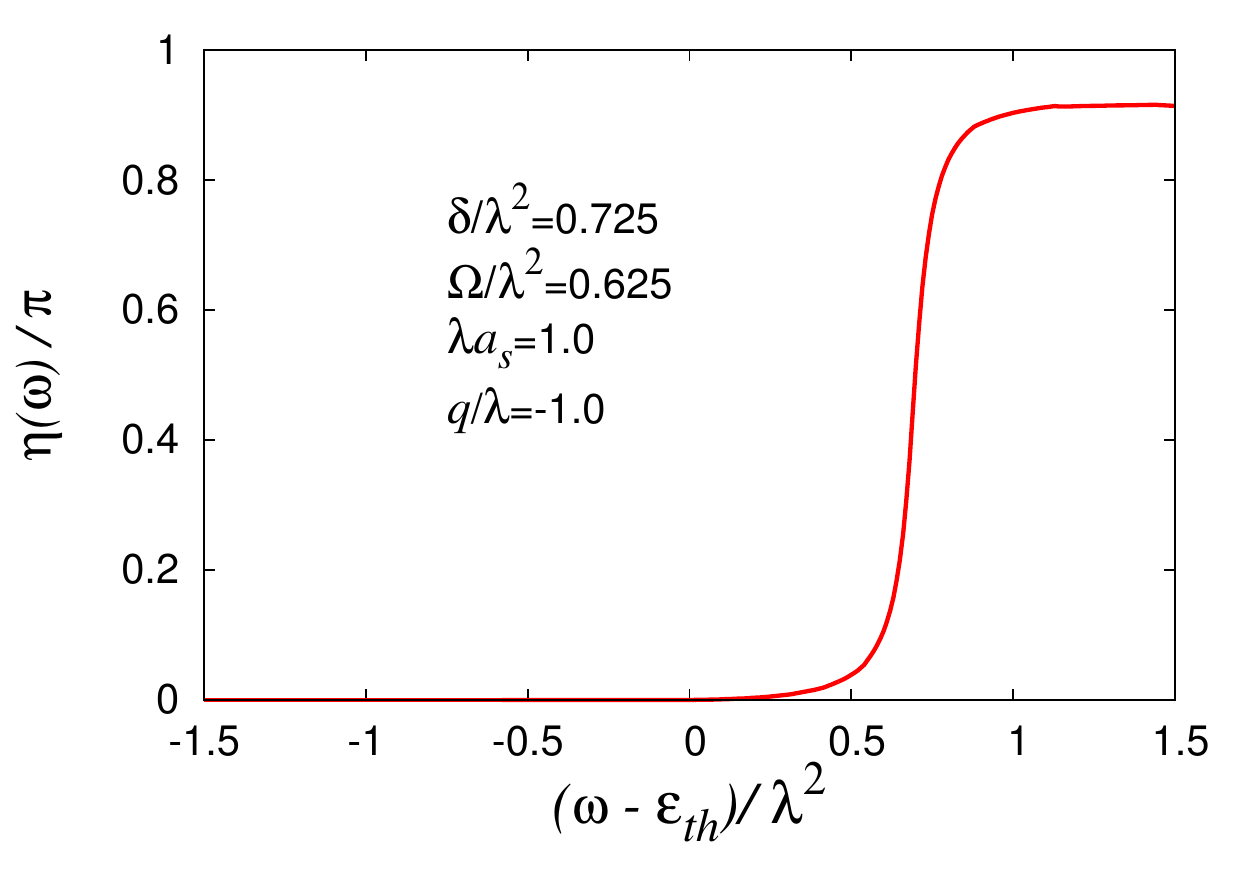}}
\caption{{\bf Scattering phase shift:} For $q \notin (q_1, q_2)$ (see \Fig{fig:Omega}), a resonance feature appears in the scattering continuum with a concomitant large increase in the scattering phase shift $\eta$.}
\mylabel{fig:PhaseShift}
\end{figure}

\medskip

\noindent
{\sc{Many-Body System:}} We now consider a system of fermions at a finite density $\rho_0 = \frac{\kf^3}{3 \pi^2} $ with  an associated energy scale  $E_F = \frac{\kf^2}{2}$. As was learnt from the analysis of the two body problem, interesting physics occurs when both $\detune$ and $\Omega$, along with $\lambda$, are non-zero, and we shall only consider such a case here.

Noting that the bound state is deeper at a finite $q$ (see \Fig{fig:Omega}), we consider many-body states corresponding to a ``flowing superfluid''. This can be done by constructing a mean field theory with $\upsilon \mean{S^\dagger(\bq)} = \frac{V \Delta}{\sqrt{2}}$. After some algebra, we obtain the following mean field Hamiltonian
\beq\mylabel{eqn:HMF}
\begin{split}
{\cal H}_{MF} & = \halfsum{\bk} \Psi^\dagger(\bq,\bk) \dH(\bq,\bk) \Psi(\bq,\bk) \\ 
& + \halfsum{{\bk}} \mbox{tr}\left(H(\frac{\bq}{2} - \bk) - \mu \bOne \right) \,\,-  \frac{V \Delta^2}{\upsilon},
\end{split}
\eeq
where the symbol $\halfsum{}$ stands summation over all $\bk$ with $k_x \ge 0$, $\Psi(\bk) = (c_{(\frac{\bq}{2} + \bk) \uparrow} \,\, c_{(\frac{\bq}{2} + \bk) \downarrow} \,\, c^\dagger_{(\frac{\bq}{2} - \bk) \uparrow} \,\, c^\dagger_{(\frac{\bq}{2} - \bk) \downarrow} )$, 
\beq\mylabel{eqn:Hdef}
\dH(\bq,\bk) = \left( \begin{array}{cc}
H(\frac{\bq}{2} + \bk) -\mu \bOne & i \Delta  \bsig_y \\
-i\Delta  \bsig_y & - \left(H(\frac{\bq}{2} - \bk) -\mu \bOne \right)
\end{array}
\right),
\eeq
where $H(\bk)$ is the matrix defined by \eqn{eqn:RSO}, $\bOne$ and $\bsig_y$ are the identity and Pauli-$y$ matrix, all three of which are $2\times2$ matrices, and $\mu$ is the chemical potential. The thermodynamic potential $P$  at a temperature $T$ is 
\beq \mylabel{eqn:Free}
\begin{split}
V P(T,\mu) &= -T \halfsum{{\bk} n} \ln(1 + e^{-E_n(\bq,\bk)/T})  \\
& + \halfsum{{\bk} \sigma} \mbox{tr}\left(H(\frac{\bq}{2} - \bk) - \mu \bOne \right) \,\,-  \frac{V \Delta^2}{\upsilon},
\end{split}
\eeq
where $n$ runs over the four eigenvalues of $\dH(\bq,\bk)$. Equilibrium values of $\Delta$ and $\mu$ are obtained by the gap and number equations derived from \eqn{eqn:Free}.

Let us discuss the results of the calculations. The first point to be noted is that in the presence of the Zeeman field, any attractive interaction does not produce a superfluid state. A non-vanishing critical scattering length $\asc$ (which is $\bq$ dependent)  is necessary to induce pairing.  This is analogous to what is known in the usual Fermi gases (without spin-orbit coupling) with imbalance. The physics of this, again, owes to the singlet density of states.

\Fig{fig:MB} shows the dependence of $\Delta$ on $q$ ($\bq = q \be_x$). The magnitude of $q$ at which $\Delta$ vanishes  (related to the Landau critical velocity) is {\em different} for flow in the $+$ direction and $-$ direction. In fact, the critical velocity for flow in the $+$ direction is about $5\%$ larger than for flow in the $-$ direction. This asymmetric  dependence of $\Delta$ on $q$ is already seen near $q=0$ (see inset of \Fig{fig:MB}), where it is clearly seen that $\Delta$ peaks at a {\em non-zero} value of $q$ which is an instance of {\em flow enhanced pairing}. We emphasize again that this phenomenon is absent when $\detune\ne 0$ and $\Omega=0$. 

\begin{figure}
\centerline{\includegraphics[width=\myfigwidth]{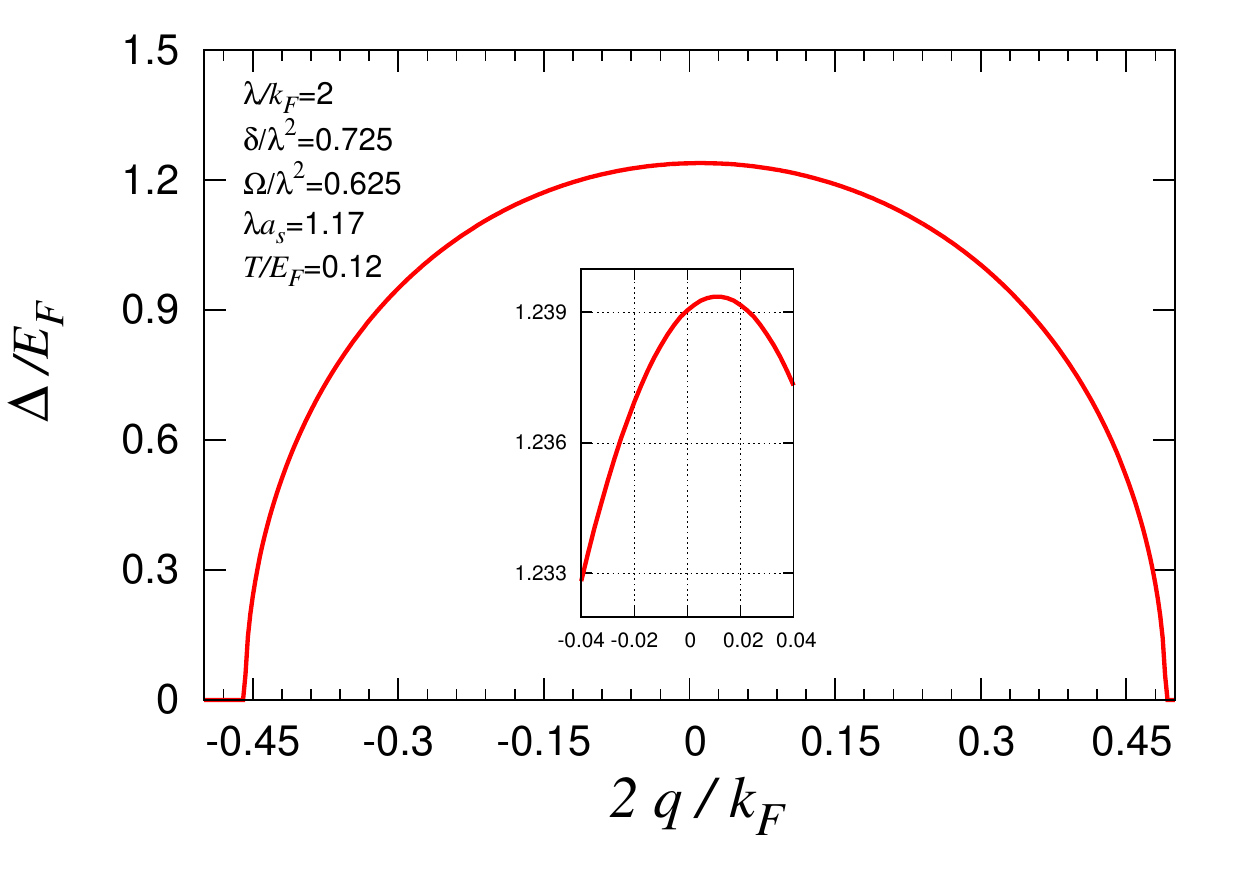}}
\caption{{\bf Dependence of pairing amplitude on $q$:} Landau critical momentum ($q$ at which $\Delta$ vanishes) for flow $+$ direction is larger than in $-$ direction. Inset shows flow enhanced pairing where $\Delta$ attains a maximum value at a nonzero $q$. }
\mylabel{fig:MB}
\end{figure}

The results just discussed may produce an impression that the asymmetric nature of the critical velocity on the direction of flow and the flow enhanced pairing, albeit clearly demonstrated, is a small effect that might be difficult to detect in experiments. While this is true at large densities, the effect is strongly enhanced at lower densities as we show by the analysis below. Consider a system where $\detune/\lambda^2 $ and $\Omega/\lambda^2$ are kept fixed, and $k_F/\lambda \ll 1$, i.~e., a low density system. Analysis of $q$ dependent pairing in this case is simplified by transforming the mean-field Hamiltonian (\eqn{eqn:HMF}) to the helicity basis
\beq
\begin{split}
{\cal H}_{MF} & = \halfsum{\bk \alpha} \left[\xi_\alpha(\frac{\bq}{2} + \bk) c^\dagger_{(\frac{\bq}{2} + \bk)\alpha} c_{(\frac{\bq}{2} + \bk)\alpha} \right. \\
& \left. - \xi_\alpha(\frac{\bq}{2} - \bk) c_{(\frac{\bq}{2} - \bk)\alpha} c^\dagger_{(\frac{\bq}{2} - \bk)\alpha}\right] + \halfsum{\bk \alpha}\xi_{\alpha}(\frac{\bq}{2} + \bk) \\
& + \Delta \halfsum{\bk \alpha \beta} A_{\alpha \beta}(\bq,\bk)  c^\dagger_{(\frac{\bq}{2} + \bk)\alpha}  c^\dagger_{(\frac{\bq}{2} - \bk)\beta} + \mbox{h.c.}  -  \frac{V \Delta^2}{\upsilon}.
\end{split}
\eeq
where $\xi_\alpha(\bk) = \varepsilon_\alpha(\bk) - \mu$.
Setting $T=0$, we focus on the regime of $q$ in $(q_1,q_2)$ (see \Fig{fig:Omega}). For a scattering length such that $\frac{1}{\as} > \frac{1}{\asc}$, in the limit of $k_F/\lambda \ll 1$, we have $|\Delta| \ll |\mu|$. Analysing the gap equation in this limit via a second-order perturbation theory in $\Delta$ gives
\beq\mylabel{eqn:BEC}
\begin{split}
\frac{1}{\upsilon} & = -\frac{1}{V} \halfsum{\bk \alpha \beta} \frac{|A_{\alpha \beta}(\bq,\bk)|^2}{\xi_\alpha(\frac{\bq}{2} +\bk) + \xi_\beta(\frac{\bq}{2} -\bk)}   = \frac{1}{V} \halfsum{\bk \alpha \beta} \frac{|A_{\alpha \beta}(\bq,\bk)|^2}{2 \mu - \varepsilon_{\alpha \beta}(\bq,\bk)} 
\end{split}
\eeq
which is the secular equation (\eqn{eqn:SecularEquation}) corresponding to the two body problem. What we have demonstrated is that in the regime $k_F/\lambda \ll 1$, and $q \in (q_1,q_2)$, the system is a BEC of the tightly bound pairs described by the bound state shown in \Fig{fig:Omega}. What is even more interesting is that, this BEC becomes a ``more robust'' superfluid with increasing $q$ starting from $q_1$; indeed, there is a $q_m \in (q_1,q_2)$ where the binding is the largest (see \Fig{fig:Omega}), clearly demonstrating that flow promotes superfluid pairing in this regime. It will be interesting to explore ways to demonstrate this in experiments.

The analysis above bring out another very interesting aspect of
this system. What happens in the regime where $k_F/\lambda \ll
1$ and $q \notin (q_1,q_2)$? The system will be a normal
fluid. What is remarkable is that this {\em normal fluid will
  be strongly interacting} due to the presence of the resonance
like feature in the scattering continuum. This state
should be accessible in the current experiments. The idea would
be to tune the parameters to a regime where $q_1 > 0$ so that
even at $q=0$ the system is in this strongly interacting normal
state with the Fermi energy tuned to the position of the resonance to produce strong interactions. In fact, the {\em broad resonance} of Li at 830 Gauss
can be utilized to study such strongly interacting Fermi
liquids, which are likely to show strong pseudogap features. An
all important question is weather such a state is unstable
towards other orders such as magnetism. Our findings,
therefore, provide an interesting direction of future research.

In summary, we have demonstrated that spin-orbit coupled systems, with detuning {\em and} Zeeman fields,  provide scope for study of a variety of new phenomena of interest to condensed matter physics. We hope that this work stimulates further experiments.

\noindent
{\bf Acknowledgement:} Support for this work by DAE (SRC grant) and DST, India is gratefully acknowledged. The author thanks, with pleasure, Jayantha Vyasanakere for discussions and suggestions. The author thanks Han Pu for intimating him about related work being carried out at Rice University through an unpublished note. Comments from the ETH Lithum team (T.~Esslinger, J.-P.~Brantut, S.~Krinner,  J.~Meineke, D.~Stadler) are acknowledged.

\bibliography{bibliography_flowsf}

\end{document}